\documentclass[twocolumn,showpacs,aps,pre]{revtex4}
\usepackage{graphicx}
\usepackage{dcolumn}
\usepackage{bm}

\begin{document}

\title{Lack of Pinning for Rigid Sliding Monolayers in Microbalance
Experiments} 
\author{J. B. Sokoloff$^*$ and I. Webman$^{**+}$}
\affiliation{$^*$Physics Department and Center for Interdisciplinary Research on
Complex Systems, Northeastern University, Boston, MA 02115; $^{**}$Physics 
Department, Bar-Ilan University, Ramat-Gan, Israel, $^+$Deceased.}
\date{\today }

\begin{abstract}

Recent work on  the dynamics of monolayers on a metallic substrate  
attached to a quartz oscillator has provided  interesting data on kinetic 
friction at the microscopic level. Sliding of the film relative to the substrate is 
often observed  even in situations in which theory seems to predict that 
the film should be 
pinned by substrate imperfections.  In this letter we propose, in order to 
attempt to resolve this issue, that if the defect potentials have a range of 
a little more than an atomic spacing, the net forces on the film due to the defects 
are likely to be quite small due to cancellations.   
\end{abstract}


\maketitle

Quartz crystal microbalance (QCM) studies of mono-layers of molecules 
on metallic substrates[1] provide detailed information about friction 
at the atomic level. The QCM consists of a  quartz oscillator of frequency 
$\sim 10^6 Hz $. Monolayer films adsorbed on the metallic surface of the quartz 
crystal oscillator are found to slip with respect to the surface during a period 
of the oscillator. Small changes observed in the frequency and in the associated 
Q factor allow us to gauge the 
friction between mono-layer  and substrate.  The amount of 
dissipation generated under most conditions implies that sliding motion of more 
than a lattice constant occurs during a period of sliding despite  
interactions that pin them together.  

 For film atom mass $m\approx 10^{-22}g$, 
appropriate for xenon atoms, QCM frequency $\omega\approx 10^7 rad/s$ and 
amplitude $A\approx 100 A^o$, which are appropriate parameters for the QCM 
experiment, the inertial force $m\omega^2 A$ is only about $10^{-14}$ dyn 
per film atom. Calculations 
of pinning of a monolayer film by defects, based on both perturbation 
theory[2] and on molecular dynamic simulations[3] seem to 
imply that this inertial force per atom due to oscillations of the substrate 
should not be sufficiently strong to overcome the pinning 
resulting from the defects. This is in fact what was found in 
experimental studies by Taborek[4] to attempt to reproduce 
the striking temperature dependence of the sliding friction found by 
Dayo and Krim for a nitrogen film sliding on a lead substrate, as 
the temperature dropped below the superconducting transition temperature 
of the lead[5]. It has been established [6], however, that the substrates 
used in this experiment were much rougher than those used in Ref. 5. 
More recent microbalance experiments done at lower oscillator 
amplitudes than those that were used in Ref. 1 find that the 
slip-time drops to zero below a certain amplitude [7], suggesting that 
the sliding film has become pinned. Since for such small amplitudes, 
the films slide by less than a lattice spacing, however, it is not clear 
that such experiments really describe sliding films. 
Recent studies by Mason et. al.[6], for this system find that the degree of 
sliding motion can be highly temperature dependent at low 
temperatures. Previous molecular dynamics studies do seem to show some thermal 
activation of the sliding but the poor statistics at slow sliding 
speeds prevents one from determining whether this is a finite size 
effect or whether it will survive for an infinite film[2], 
considering that the films in Ref. 3 were only about 13 lattice spacings 
across. The relevant 
interaction strengths indicate a {\it weak pinning} situation  characterized by a  
Larkin length $L_c$, large compared to a lattice spacing, 
within which elastic deformation due to pinning interactions is small[8]. 
Despite the fact that the defect potential strengths are comparable or even 
smaller than $k_B T$ where $k_B$ is Boltzmann's constant and T is the 
temperature, one might expect that since the film is rigid over a length $L_c$, 
one must thermally depin a whole Larkin domain in order to achieve thermally 
activated depinning of the film, which require values of $k_B T$ much greater 
than a typical defect potential strength. 
In the usual treatment of an elastic 
medium interacting with a disordered potential \cite{larkin}, the range 
of the defect potential is assumed to be sufficiently small compared to 
a film lattice spacing, so that it can only interact with one atom at a time. 
In contrast, pinning defects on the substrate typically extend  
over one or more atomic lengths.  It will be 
argued here that for defect potentials, typical of surface defects, that 
have a long enough range to interact with two or more atoms the forces on 
these atoms tend to cancel each other, reducing the force on the film to 
a value much smaller than the maximum force that a single defect can exert 
on a single film atom. 

The fact that the Larkin length is many lattice spacings long 
indicates that the film is sufficiently stiff that it distorts over 
distances long compared to the mean defect potential spacing, and hence 
certainly long compared to the defect potential range. It will now be 
demonstrated that because of the film stiffness, even for defect potentials 
with a range of a little more than a film lattice spacing the forces 
exerted on the film atoms tend to cancel to a great degree. 
(Incidentally, the simulations reported in Ref. 3 used a defect potential 
of much shorter range.)  Consider the potential energy of a rigid film of 
atoms interacting with a single defect potential v({\bf R}+{\bf r}), 
$$\sum_{\bf R} v({\bf R}+{\bf r}) \eqno (1)$$ 
where  {\bf r} represents a displacement of the film relative to the 
potential and {\bf R} is an atomic position in the periodic lattice 
of the film. Since this quantity is a periodic function of {\bf r} 
with the periodicity of the film lattice, the above summation can 
be expressed as \cite {ashcroft}  
$$\sum_{\bf R} v({\bf R})=\sum_{\bf G} \bar{v} ({\bf G})e^{i{\bf G}\cdot {\bf r}}
\eqno (2)$$
where the Fourier transform of v({\bf r}), $\bar{v} ({\bf G})=
\Omega^{-1}\int_{\Omega} d^2 r e^{-i{\bf G}\cdot {\bf r}}v({\bf G})$, where 
the integral is taken over the film unit cell area $\Omega$ and 
{\bf G} is a reciprocal lattice vector of the film. Since the form 
of the defect potential for local defects such as vacancies or interstitials 
is not known, let us study a couple of simple potentials which drop off 
reasonably rapidly at large distances, in order to develop a picture 
of what one expects for defect potentials with a range greater than 
a lattice spacing. For a Gaussian defect potential, 
$v({\bf r})=-V_0 e^{-(r/b)^2},$ where b is the range parameter, we 
obtain, using Eq. (2)
$$\sum_{\bf R} v({\bf R}+{\bf r})=$$
$$-V_0 \pi (b/a)^2 
[\sum_{\bf G} e^{-(G^2 b^2/2)} e^{i{\bf G}\cdot {\bf r}}]. \eqno (3)$$
For example, for a range parameter b=1.23a, where a is a film lattice constant, 
for a square lattice, $e^{-(G^2 b^2/2)}=3.28\times 10^{-7}$, whereas for 
b=0.83, it is only 0.0011. (Similar calculations on a triangular lattice give comparable results.) Thus, even for a range b as short as 1.23a, the {\bf r} dependent part 
of the interaction with the lattice (i.e., the terms in the expression in 
Eq. (2) with {\bf G} not equal to zero) is much less than 
$V_0$. For v(x)=$-V_0 [(1+(x/a)^2)(1+(y/a)^2)]^{-1}$, $\bar{v} ({\bf G})$ 
on the right hand side of Eq. (2) becomes proportional to $\pi b e^{-(|G_x|+|G_y|)b}$, 
whose largest value for a square lattice is 0.000440 for b=1.23a and 0.00543 
for b=0.83a. Thus, for both of these defect potentials, the interaction of 
the film with the potential is smaller than $V_0$, and decreases rapidly with 
increasing b. 


At any temperature, there will be lattice vibrations. At a given 
instant of time, a lattice vibration of wavevector {\bf q} will add a term 
${\bf A} cos({\bf q}\cdot ({\bf r}+{\bf R}))$ to {\bf R} in Eqs. (1,3), where {\bf A} 
is the amplitude of the vibrational mode, assumed to be much smaller than 
a lattice constant a. The potential energy term in the summation in Eq. (1) 
is still periodic, and hence Eq. (2) is still valid. If we expand the 
potential to first order in {\bf A}, we find that $\bar{v} ({\bf G})$ for 
the Gaussian potential now contains a term $e^{-|{\bf G}-{\bf q}|^2 b^2/2}$, 
which can be larger than  $e^{-G^2 b^2/2}$ by a sufficient amount 
to make the former term dominate over the latter for values of q 
which are comparable in magnitude to the smallest values of G, even 
when it is multiplied by {\bf A}, which is much less in magnitude than a. 
As the latter term is proportional to 
{\bf A} which oscillates in time for a lattice vibration, this term 
produces an oscillating term in the potential energy, and hence, a 
force on the film, which is
larger in magnitude than the potential energy and force on the film 
that would occur if there were no vibrations. Thus at even fairly 
low temperatures, the lattice vibrations can produce oscillating 
forces acting on the film, which dominate over the static pinning force 
on the film, implying that even at low temperatures, the film will 
not be pinned by the defects because the film is pushed out of its 
total potential minimum by this force. 

In order to illustrate this effect, consider a rigid two dimensional 
square lattice of atoms of lattice constant a interacting with a Gaussian 
potential $-V_0 e^{-r^2/b^2}$, were r is the distance from the center of 
the potential. We choose the range parameter of the potential, 
b=1.236067978a. For this choice of parameters the potential range is 
about one and a half lattice spacings. The lattice 
is slid a distance s and the total force due to this potential acting on 
it is plotted as a function of s (the lower curve in Fig. 3a). The 
position of an atom in the lattice is then displaced by an amount 
$0.01acos((2\pi/\lambda)x_n)$ in the x-direction,  where the position 
of a film atom is $(x_n,y_m)=(na,ma)$ where n and m are integers,   
to simulate the effect of short wavelength 
vibrations. Here $\lambda$ is taken to be equal to b/3. The total force acting 
on the lattice is again plotted as a function of s for this case in Fig. 3a 
(the higher curve). The total force acting on the lattice was found to be three 
orders of magnitude larger for the modulated than for the unmodulated lattice. 
In this example, a static modulation was used. For a modulation due to 
actual thermally activated phonons, the modulation oscillates quite rapidly 
in time, resulting in a rapidly oscillating force, which 
overcomes any weak static pinning force due to the average interaction of 
the defect potential with the lattice. Thus even if the inertial force 
is not large enough to depin a film without lattice vibrations, this 
oscillating force may depin it at sufficiently high temperature. 
Even the zero point oscillations 
of the lattice occurring at low temperatures are larger than the 
amplitude of the vibrations assumed here \cite {ziman}.
For comparison, in Fig. 3b, we present results of the same model 
calculations illustrated 
in Fig. 3a but with the value of b 67.5\% of the value used in 3a. It is 
quite clear that the effect illustrated in Fig. 3a has almost completely 
disappeared. 
\begin{figure}[tbp]

\center{
\includegraphics [angle=0,width=2.05in]{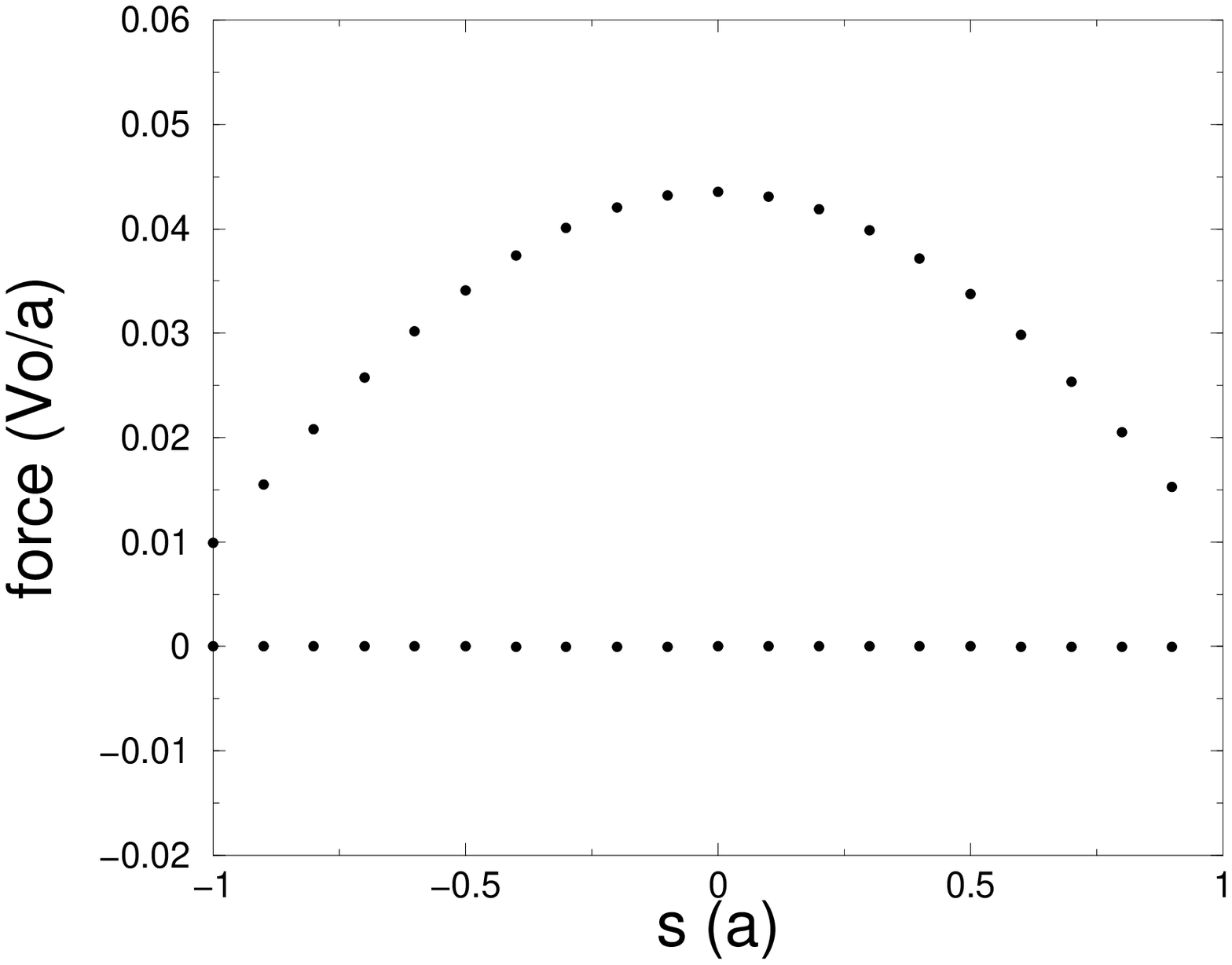} 
\vskip -0.3cm
\includegraphics [angle=0,width=2.05in]{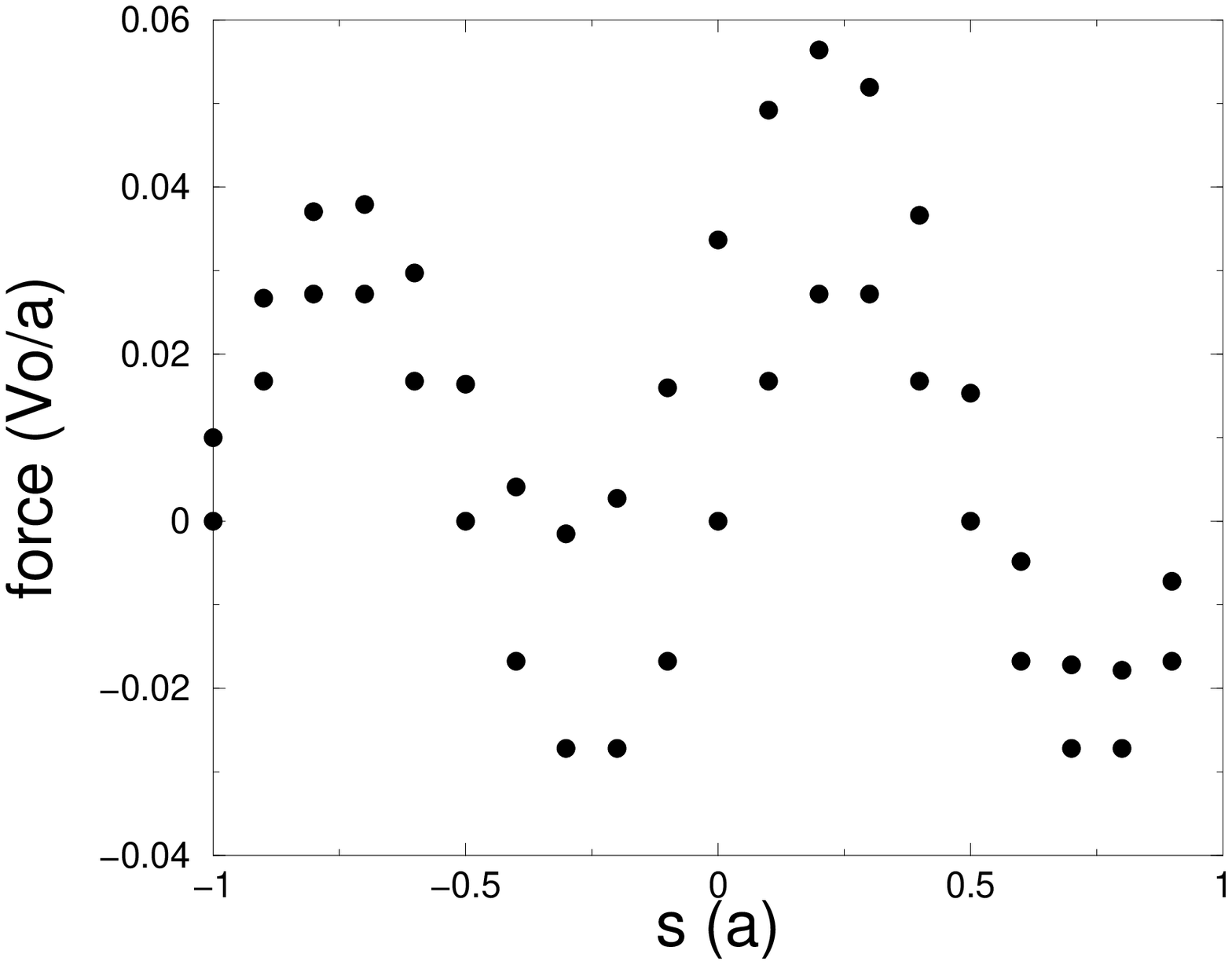}
 \vskip -0.3cm}
 \vskip -0.3cm

\caption{a. The total force due to a Gaussian potential acting on a rigid 
lattice  of atoms of spacing a is plotted as a function of the displacement 
of the chain s in the lower nearly flat curve for b=1.236067978a. The total 
force acting on a modulated chain is shown in the higher curve. 
b. The total force due to the Gaussian potential for both modulated and 
unmodulated films like those in a is shown for b=0.8343349a.}
\label{fig1}
\end{figure}

In conclusion, we have proposed a mechanism which may explain why local defects, 
which must certainly be present on even the smoothest surfaces, might not 
prevent a stiff monolayer film from sliding in a quartz microbalance experiment, if 
the defect potential has a range greater than a film lattice spacing. 
Our proposed mechanism is supported by a simple model calculation, based on the fact 
that each defect can interact with several film atoms at a time. As a consequence, 
the forces on the film atoms tend to cancel. Lattice vibrations are able 
to provide an oscillating force which is considerably larger than the net 
static force due to 
the defect potential acting on the film. This can lead to a high 
degree of thermal activation of the 
sliding of the film.  
\section*{Acknowledgments}
J.B. Sokoloff wishes to thank the Department of Energy 
(Grant DE-FG02-96ER45585).

\end{document}